\documentclass[sigconf, screen]{acmart}

\usepackage{geometry}
\usepackage{tabularx}
\usepackage{dcolumn} %
\newcolumntype{d}[1]{D{.}{.}{#1}}
\usepackage{svg}
\graphicspath{{figures/}}
\usepackage[capitalize]{cleveref}

\AtBeginDocument{%
  }

\copyrightyear{2025} 
\acmYear{2025} 
\setcopyright{rightsretained}
\acmConference[CHI EA '25]{Extended Abstracts of the CHI Conference on Human Factors in Computing Systems}{April 26-May 1, 2025}{Yokohama, Japan}
\acmBooktitle{Extended Abstracts of the CHI Conference on Human Factors in Computing Systems (CHI EA '25), April 26-May 1, 2025, Yokohama, Japan}\acmDOI{10.1145/3706599.3720243}
\acmISBN{979-8-4007-1395-8/2025/04}

\begin{document}

\title[Composable Prompting Workspaces for Exploration and Iteration]{Composable Prompting Workspaces for Creative Writing: Exploration and Iteration Using Dynamic Widgets}

\settopmatter{authorsperrow=4}

\author{Rifat Mehreen Amin}
\orcid{0000-0003-4279-7778}
\affiliation{%
  \institution{LMU Munich}
  \city{Munich}
  \country{Germany}
}
\email{rifat.amin@ifi.lmu.de}

\author{Oliver Hans Kühle}
\orcid{0009-0000-3581-2111}
\affiliation{%
  \institution{LMU Munich}
  \city{Munich}
  \country{Germany}
}
\email{o.kuehle@campus.lmu.de}

\author{Daniel Buschek}
\orcid{0000-0002-0013-715X}
\affiliation{%
 \institution{University of Bayreuth}
  \city{Bayreuth}
  \country{Germany}}
\email{daniel.buschek@uni-bayreuth.de}

\author{Andreas Butz}
\orcid{0000-0002-9007-9888}
\affiliation{%
 \institution{LMU Munich}
  \city{Munich}
  \country{Germany}}
\email{andreas.butz@ifi.lmu.de}

\renewcommand{\shortauthors}{Amin et al.}

\begin{abstract}
Generative AI models offer many possibilities for text creation and transformation. Current graphical user interfaces (GUIs) for prompting them lack support for iterative exploration, as they do not represent prompts as actionable interface objects. We propose the concept of a composable prompting canvas for text exploration and iteration using dynamic widgets. Users generate widgets through system suggestions, prompting, or manually to capture task-relevant facets that affect the generated text. In a comparative study with a baseline (conversational UI), 18 participants worked on two writing tasks, creating diverse prompting environments with custom widgets and spatial layouts. They reported having more control over the generated text and preferred our system over the baseline. Our design significantly outperformed the baseline on the Creativity Support Index, and participants felt the results were worth the effort. This work highlights the need for GUIs that support user-driven customization and (re-)structuring to increase both the flexibility and efficiency of prompting.
\end{abstract}

\ccsdesc[500]{Human-centered computing~Interactive systems and tools}
\ccsdesc[500]{Human-centered computing~Graphical user interfaces}

\keywords{Dynamic UI, Prompting, LLM, human-AI co-creation, creativity support}

\begin{teaserfigure}
  \includegraphics[width=\textwidth, trim={0cm 0cm 0cm 0cm},clip]{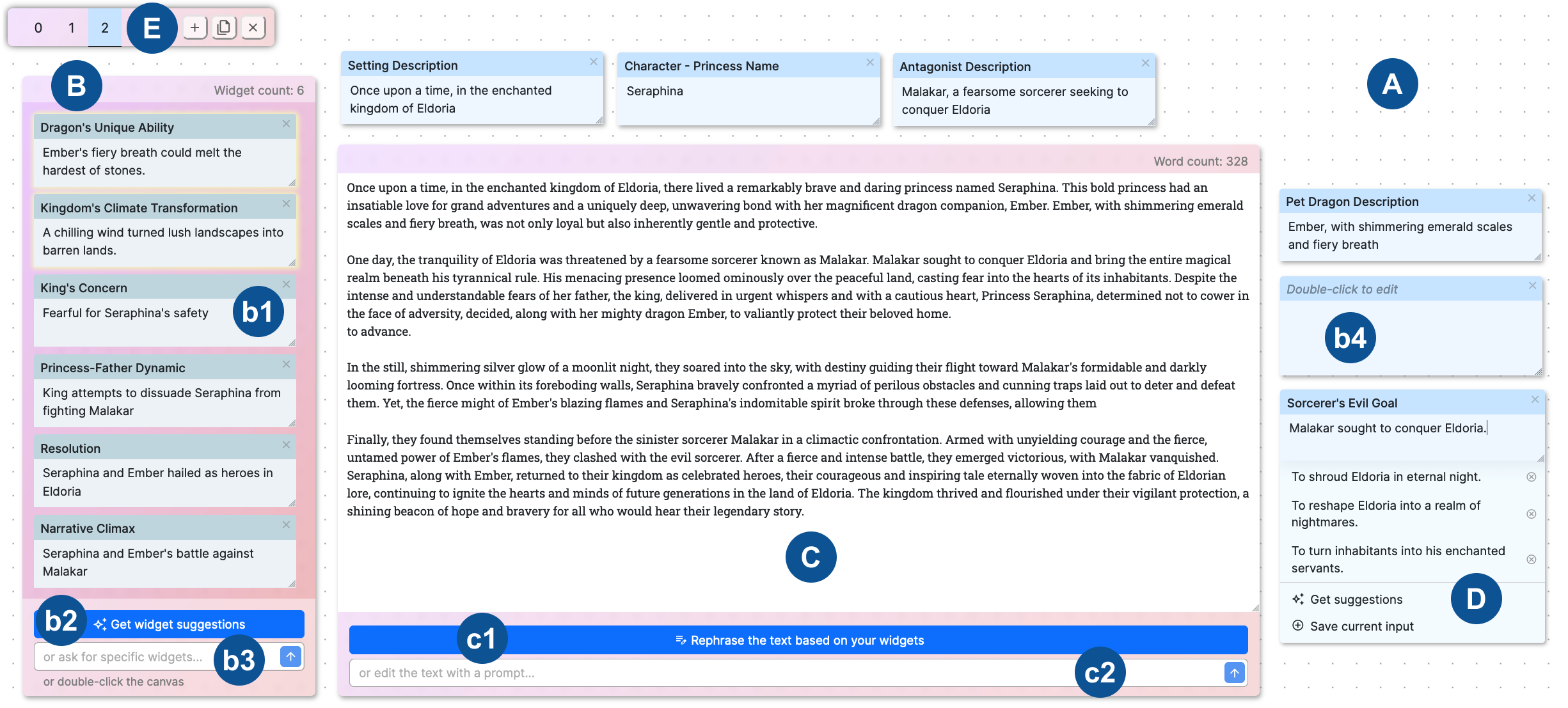}
  \caption{User interface of PromptCanvas. (A) Canvas-like workspace where users can place and organize widgets to create a customized environment. (B) Widget panel. (b1) Example of widgets created through system suggestions. (b2) Button to get widget suggestions from the system. (b3) Field for entering prompts to create multiple widgets of a specific theme. (b4) Example of an empty widget created by double-clicking at empty space. (C) Text editor and output text generations from the prompt. (c1) Button to rephrase the text based on the widgets on the canvas (light blue). (c2) Field to provide prompts for generating text. (D) Example of an opened widget with suggested values for customization. In each widget, users can request more such suggestions and save current input for refinement and iteration. (E) Menu bar for creating, copying, or deleting a canvas.}
  \Description{The image shows a screenshot of PromptCanvas. It shows all the elements of our UI in one image.}
  \label{fig:teaser}
\end{teaserfigure}

\maketitle

\section{Introduction}

Advancements in generative AI models have revolutionized text interaction, offering powerful tools for creating and exploring text~\cite{Suh_Chen_Min_Li_Xia_2024, 10.1145/3613904.3642105, reif2024automatic}. These may enhance creative expression by providing users with novel ways of generating text and interacting with it. However, their potential is often constrained by the limitations of existing graphical user interfaces (GUIs). The primary limitation of current GUIs for prompting generative AI models lies in their inability to support iterative exploration and customization. These GUIs present prompts as static text fields, restricting users to a linear interaction paradigm~\cite{Jiang_Rayan_Dow_Xia_2023, goodwin2015professional}. For writers, this approach can lead to what \citet{kreminski2024dearth} refers to as ``dearth of the author'' -- a condition in which users become disengaged from the creative process and produce text that lacks expressive intent. This lack of interactivity and flexibility hinders users' ability to leverage generative AI capabilities creatively. Users may find it challenging to achieve their desired outcomes without the ability to dynamically manipulate prompts, create personalized workflows, or easily explore a wide range of variations. Additionally, the metacognitive demands placed on users by generative AI tools further exacerbate these challenges~\cite{metacognition}.

To address these limitations, we introduce a novel approach to enhance prompting in creative writing, inspired by the concept of dynamic widgets, introduced by \citet{dynavis_2024} for information visualization. We bring dynamic widgets to writing: Our system, PromptCanvas, empowers users to create custom GUIs tailored to their writing needs. Concretely, PromptCanvas transforms prompts into actionable and persistent interface objects by allowing users to dynamically arrange and customize widgets on a canvas. These widgets offer interactive elements based on the context of the prompt, providing flexibility and control over customizable, relevant aspects of the generated text. This allows users to create personalized prompting environments that reflect their unique workflows and creative styles, facilitating iterative refinement of their own draft or AI-generated text. Beyond customizability, dynamic widgets can support metacognition by assisting in task decomposition 
and promoting a more structured, iterative use of generative AI.

Our study reveals that dynamic widgets improve user experience by enhancing control over text generation, reducing cognitive load, enabling iterative exploration, and supporting diverse prompts. These findings highlight the value of customizable writing tools. By prioritizing flexible, user-driven customization with dynamic widgets, we advance human-AI interaction and promote more creative interaction with generative AI.

In summary, this research contributes to human-AI interaction in writing by investigating the following research questions: 

\begin{itemize}
        \item [\textbf{RQ1}] How can writing tools be designed with dynamic widgets to improve user interaction and creativity and provide greater control over the generated content?
        \item [\textbf{RQ2}] Do dynamic widgets for iterative and structured prompting improve creativity support compared to an existing conversational UI?
    \item [\textbf{RQ3}] Do dynamic widgets help in reducing cognitive load in creative writing tasks?
\end{itemize}

\section{Background and Related Work}
\subsection{Dynamic and Adaptive UIs in Creative Workflows}
Early work by \citet{ahlberg1994visual} highlighted the benefits of tightly coupling user inputs with outputs, fostering engagement and immediate feedback. However, in hindsight these systems were limited by static UI elements. Recent developments, such as FrameKit~\cite{FrameKit}, address this by creating adaptive UIs that adjust to user context and interaction patterns, enhancing user experience \cite{findlaterDesignSpaceEvaluation2009, todiAdaptingUserInterfaces2021, KHAMAJ2024164}. Moreover, the principles of \textit{reification, polymorphism, and reuse}~\cite{beaudouin-lafonReificationPolymorphismReuse2000} introduced foundational concepts for efficient, user-centered interfaces, making abstract operations tangible, tools adaptive, and outputs reusable. Modern systems like Eviza~\cite{Eviza}, DynaVis~\cite{dynavis_2024}, and Bolt~\cite{srinivasan2023bolt} extend these ideas with natural language inputs and dynamic widgets for data visualization and modification. Widgets simplify complex tasks, as seen in Bespoke~\cite{Bespoke}, which generates GUIs from command-line inputs, and ProvenanceWidgets~\cite{provenancewidgets}, which tracks and visualizes user interactions. Collectively, these advances enable dynamic exploration, adjustment, and refinement, fostering creativity and enhancing user productivity.

\subsection{Intelligent and Interactive UIs for Prompting}
Recent advancements in generative AI have led to interactive systems that allow users to create content  through natural language inputs (prompts). A recent line of work on such systems explores user agency via direct manipulation, such as in \textit{Spellburst}~\cite{Spellburst}, which uses a node-based interface for semantic programming and exploring variations, and \textit{DirectGPT}~\cite{masson}, which allows users to modify generated content via graphical objects and direct controls. %
\textit{Low-code LLM}~\cite{caiLowcodeLLMGraphical2024} enables prompt creation with drag-and-drop functionality, while \textit{Textoshop}~\cite{massonTextoshopInteractionsInspired2024} adapts graphic-editing tools for intuitive text manipulation. Personalization is also advancing, as seen in \textit{Writer-Defined AI Personas}~\cite{benharrakWriterDefinedAIPersonas2024}, which lets users create custom AI personas for tailored feedback. Systems like \textit{Sensecape}~\cite{suhSensecapeEnablingMultilevel2023}, \textit{Graphologue}~\cite{Jiang_Rayan_Dow_Xia_2023}, and \textit{Luminate}~\cite{suh2024luminate} provide interactive visualizations to structure and refine content, enabling deeper user comprehension. These systems represent a shift toward user-centered generative AI, empowering users to shape content through direct interaction, personalization, and accessible interfaces.

\subsection{Human-AI Collaborative Writing and Content Creation}
Integrating AI into creative processes has transformed writing and content creation by enhancing interaction and providing continuous feedback. Tools like those discussed by \citet{dangTextGenerationSupporting2022} support writing momentum, reducing creative block, while \citet{gilburt2024machine} highlights how AI helps overcome writer's block by reigniting stalled ideas. Generative AI is applied across domains, including code generation~\cite{austinProgramSynthesisLarge2021}, email auto-completion~\cite{choice}, comic creation~\cite{codetoon}, screenplay co-writing~\cite{mirowski2023cowriting}, argument drafting~\cite{Zhang_2023}, and academic writing~\cite{nguyen2024human}. Professional perspectives on this transformation are captured by \citet{ippolitoCreativeWritingAIPowered2022}. Challenges remain as AI becomes a co-creator. Research by \citet{control} explores interaction with prompting during writing, while \citet{metacognition} examines the metacognitive demands and opportunities of generative AI tools. Critical assessments by \citet{kreminski2024dearth} and \citet{mirowski2023cowriting} address AI’s reception in creative industries and areas for improvement. Overall, AI's integration in writing is transforming workflows, demanding new interfaces to support fruitful use.

\section{System Design}

The current design of PromptCanvas emerged through multiple iterations of planning and design sessions conducted by the authors for achieving specific design goals (DGs). These sessions involved brainstorming, prototyping, and refining the interface. 
This iterative process allowed us to explore different layouts and widget functionalities to ensure the interface supports creativity, flexibility, and ease of use. This process also implicitly answers RQ1.

\subsection{Design Goals and Proposed Design Solutions}

\begin{itemize}
    \item \textbf{DG1. Transform prompts into visible and actionable objects.} Current interfaces treat prompts as static text fields, limiting user interaction to basic input-output cycles. PromptCanvas converts prompts into dynamic widgets that represent actionable components of the text. These widgets allow users to adjust attributes, such as tone, style, or content interactively, enabling more granular and intuitive control over text generation by providing all the benefits of direct manipulation interfaces~\cite{shneiderman1983direct}.
    \item \textbf{DG2. Facilitate structured exploration and refinement.} Writing is an iterative process that requires the ability to experiment with and refine ideas systematically. PromptCanvas enables users to break down tasks into smaller components using widgets, supporting structured exploration and iterative improvement. This design ensures users can focus on individual aspects of their text while maintaining a cohesive workflow.
    \item \textbf{DG3. Promote divergent thinking and creativity.} To overcome creative blocks and encourage novel ideas, the system should support divergent thinking. Context-aware widget suggestions, parallel exploration options, and dynamic rephrasing tools help users explore multiple creative directions and ideas.
    \item \textbf{DG4. Provide a customizable and adaptable workspace.} Every user has a unique writing process, so the interface must accommodate diverse workflows. PromptCanvas, therefore, offers a flexible, infinite canvas where users can create, organize, and rearrange widgets to suit their personal preferences. This customizability allows the workspace to evolve with the user's needs.
    \item \textbf{DG5. Simplify navigation and reduce cognitive load.} The open-ended nature of the infinite canvas can be overwhelming without proper navigation aids. PromptCanvas includes features like widget panels, drag-and-drop interactions, and clear visual hierarchies to help users locate and organize their ideas efficiently, to minimize cognitive strain.
\end{itemize}

\subsection{Resulting Interface and Features}
PromptCanvas %
is built around an infinite canvas, a zoomable digital workspace that users can navigate and organize freely, in order to improve flexibility in content creation and management. This canvas is complemented by three key components: the text editor, control widgets, and the widget panel (see \autoref{fig:teaser}).

\subsubsection{Infinite Canvas}
The infinite canvas provides users with \textbf{spatial freedom} to create and organize content in a way that best suits their workflow, as shown in \autoref{fig:teaser}-(A). \textbf{Users can pan across the workspace and zoom in and out seamlessly}, enabling transitions between broad overviews and detailed views of specific elements. \textbf{This flexibility allows users to visually organize their ideas spatially} by grouping related items, layering, or arranging them hierarchically, which promotes clarity and supports systematic exploration \textbf{(DG2)}. Additionally, the open-ended layout ensures that users can customize their workspace, reflecting their unique processes and preferences \textbf{(DG4)}.

\subsubsection{Text Editor}
The text editor serves as the centerpiece of the interface, allowing users to integrate their input with system-generated suggestions fluidly, as illustrated in \autoref{fig:teaser}-(C). \textbf{Users can refine their text iteratively by rephrasing it based on active control widgets or submitting prompts} to create or modify content, making it easier to experiment with different ideas \textbf{(DG2)}. The editor also supports \textbf{incremental text generation}, displaying content dynamically as it is produced, which helps users remain engaged with the evolving output. Additionally, the \textbf{history feature enables users to revisit previous iterations}, promoting iterative improvement and exploration of alternatives \textbf{(DG3)}. These features are further complemented by real-time updates to word counts and visual indicators of changes, supporting clarity and focus during the writing process.

\subsubsection{Control Widgets}
Control widgets are dynamic, interactive tools that transform abstract text attributes into actionable and adjustable UI elements, as shown in \autoref{fig:teaser}-(b1). \textbf{Each widget provides context-aware suggestions tailored to the content in the text editor}, helping users explore multiple creative directions and overcome writer’s block \textbf{(DG3)}. These widgets allow users to \textbf{adjust text attributes like tone, style, or structure directly}, offering control over the output and turning prompts into interactive objects \textbf{(DG1)}. Their flexibility in resizing, repositioning, and customization ensures that the workspace adapts to user needs as tasks evolve \textbf{(DG4)}. Additionally, their integration with the rephrasing and text generation systems ensures a seamless workflow between ideation and implementation.

\subsubsection{Widget Panel}
The widget panel acts as the system’s central hub for generating, managing, and organizing control widgets, as shown in \autoref{fig:teaser}-(B). \textbf{Users can create widgets dynamically based on text analysis or provide specific input for guided widget creation}, making it easier to tailor tools for individual tasks \textbf{(DG1)}. The panel highlights newly generated widgets with a yellow glow and allows users to evaluate, delete, or drag them onto the canvas, ensuring only relevant widgets influence text generation. \textbf{Its visually distinct layout and dynamic updates simplify navigation and reduce cognitive load}, helping users locate and manage ideas efficiently \textbf{(DG5)}. The scrolling functionality and size adjustments aim to support projects with many widgets. %
An example scenario of how to use the system for writing a short story is provided in \cref{app: example}.

\section{User Study Design}

To answer RQ2 and RQ3, we conducted a within-subjects lab study with 18 participants between the ages of 22 and 68 years ($M = 30.44, SD = 10.45$). They had varied writing experience, including emails, letters, blogs, and stories, and used AI tools like ChatGPT, Quillbot, and Bard for writing tasks. Participants were compensated with 10€/hour. 

We evaluated PromptCanvas (``dynamic UI'') against a static conversational UI (``static UI'', \autoref{fig:static_UI}), with approval from our institution's ethics board. Participants completed two tasks using each UI, which we selected to cover different types of writing: emails (5 minutes) and short stories (10 minutes) on predefined or user-selected topics (\cref{app:writingTasks}).

Besides these writing tasks, we included a pre-study survey, an interface tutorial, and a post-study survey and semi-structured interviews for feedback. Screen and audio were recorded for analysis.  Quantitative data, such as numbers and types of prompts and widgets used, along with Creativity Support Index (CSI) and NASA Task Load Index (NASA-TLX) metrics, were analyzed statistically (Shapiro-Wilk, paired t-test, Wilcoxon signed-rank test) to determine significance. More details regarding our user study are provided in \cref{app: user study}.

\section{Results}
\subsection{How Does PromptCanvas Support Creativity and Exploration? (RQ2)}

PromptCanvas scored significantly higher than the baseline on all factors of the Creativity Support Index (CSI) (all $p < 0.03$, see \autoref{tab:CSI-within}). 
As our study did not involve collaboration, we omitted the collaboration factor following the practice from~\cite{carollCreativityFactorEvaluation2009, codetoon} to avoid confusion.

\begin{table}[t]
\caption{Creativity Support Index (CSI) Results (N=18).}
\begin{tabularx}{\linewidth}{Xd{2.2}d{2.2}d{2.2}d{2.2}d{0.3}}
\toprule
   &  \multicolumn{2}{c}{Baseline} & \multicolumn{2}{c}{PromptCanvas} & \\ %
   \cmidrule(r){2-3}\cmidrule(lr){4-5}%
   Factor  & \multicolumn{1}{c}{M}  & \multicolumn{1}{c}{SD} & \multicolumn{1}{c}{M}  & \multicolumn{1}{c}{SD} & \multicolumn{1}{c}{$p$}\\ 
   \midrule
    Enjoyment& 13.06 & 4.40 & 16.56 & 3.99 & .02\\
 Exploration& 11.78 & 5.53 & 16.83 & 3.08 &  .02\\
 Expressiveness& 10.83 & 4.91 & 14.67 & 3.76 & .02\\
 Immersion& 10.00 & 4.10 & 14.61 & 4.73 & .01\\
 Results Worth Effort& 14.17 & 4.02  & 17.61 & 2.21 & .005\\
    \midrule
    Overall CSI Score& 61.65 & 18.53 & 82.09 & 12.12 &  .005 \\ 
\bottomrule
\end{tabularx}
\Description{This table shows the Creativity Support Index (CSI) Results (N=18).}
\label{tab:CSI-within}
\end{table}

\begin{figure*}[t]
    \includegraphics[width=1\textwidth]{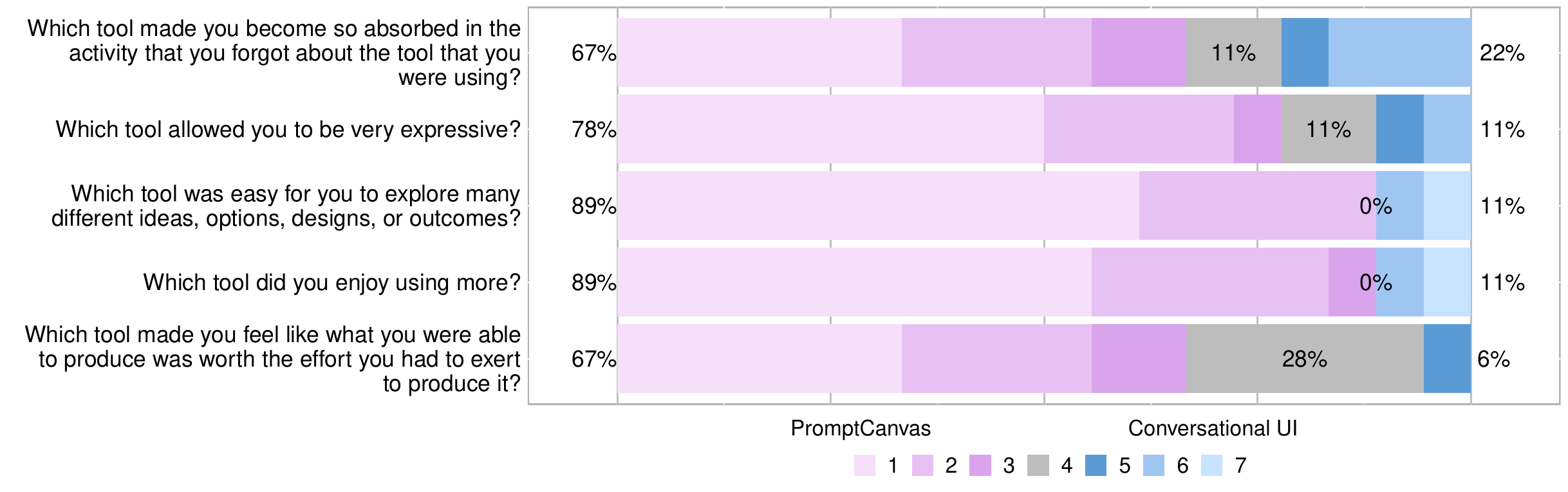}
    \caption{Self-reported creativity support scores and preferences comparing PromptCanvas and the Conversational UI (N=18).}
    \label{fig:Preference_CSI}
    \Description{This image shows participants' self-reported creativity support scores comparing the baseline and our system.}
\end{figure*}

Participants found PromptCanvas ($M=82.09, SD=12.12$) to support creativity significantly more ($p=0.005$) compared to the conversational UI
($M=61.65, SD=18.53$). The $p$-values were adjusted using the Bonferroni-Holm correction to account for multiple comparisons. In the final survey, participants directly compared the creativity support between PromptCanvas and the conversational UI (\autoref{fig:Preference_CSI}). 

67\% of participants reported that they became so absorbed in the activity that they forgot about the tool they were using. 78\% of participants chose PromptCanvas as the more expressive tool. 89\% of participants found that they explored a wider range of ideas, options, designs, or outcomes using our system compared to using the baseline.
Furthermore, 89\% of participants reported a higher level of enjoyment with PromptCanvas than with the baseline. Using PromptCanvas made users feel their efforts were most worthwhile, with 67\% feeling satisfied with what they produced relative to the effort expended.

\subsection{How Does PromptCanvas Affect the Cognitive Load in Creative Writing? (RQ3)}\label{subsec:cognitive_load}

\begin{figure*}[t]
    \centering
    \includegraphics[width=\linewidth]{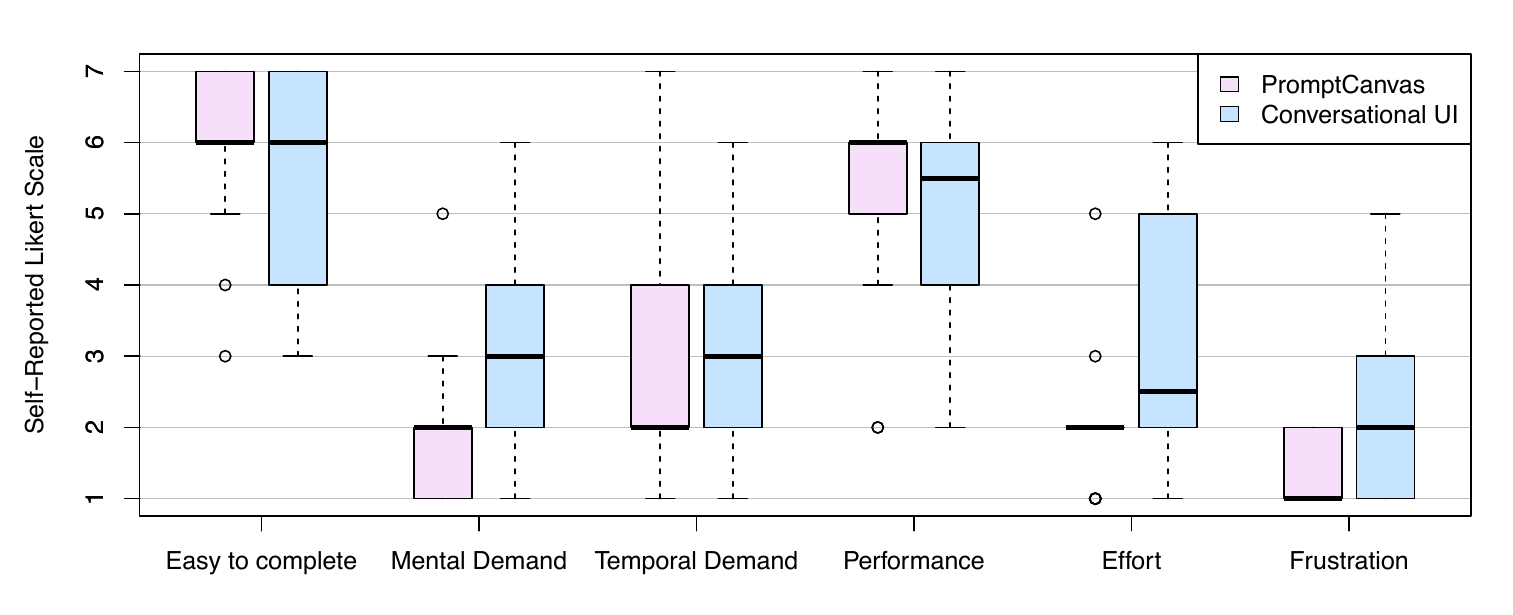}
    \caption{Self-reported NASA-TLX scores and ease-of-use ratings from participants in our lab study (N=18).}
    \label{fig:NASA-within}
    \Description{This box plot shows participants' self-reported scores for NASA TLX questions and ease of using the system.}
\end{figure*}

Participants evaluated their perceived cognitive load using the NASA-TLX scale after using each UI. Significant differences were observed in two aspects of cognitive load: mental demand and frustration. For mental demand, participants reported a significantly ($p=0.02$) lower mental demand when using the dynamic interface ($M=1.89$, $SD=1.02$, $Med=2$), compared to the static interface ($M=3.06$, $SD=1.51$, $Med=3$) (See \autoref{fig:NASA-within}). 

\begin{figure*}[t]
    \centering
    \includegraphics[width=\linewidth]{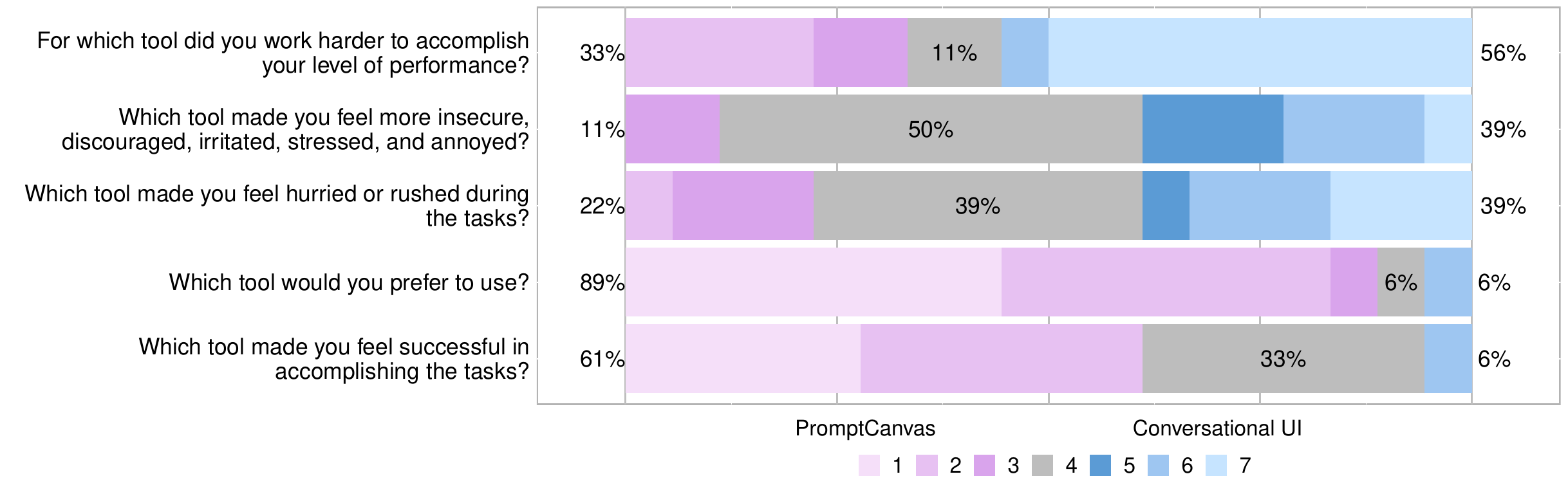}
    \caption{Self-reported cognitive load and preference scores comparing PromptCanvas and the conversational UI (N=18).}
    \label{fig:Preference_NASA}
    \Description{This image shows participants' self-reported cognitive load and preference scores that directly compare the Conversational UI and PromptCanvas.}
\end{figure*}

Frustration ratings also differed significantly ($p=0.03$), with the dynamic interface showing lower frustration ($M=1.28$, $SD=0.46$, $Med=1$) than the static interface ($M=2.17$, $SD=1.42$, $Med=2$), indicating that PromptCanvas helped reduce feelings of insecurity, irritation, and stress. However, other cognitive load aspects showed no significant differences between the two interfaces. In the final survey, participants directly compared their perceived cognitive load between PromptCanvas and the conversational UI. Results are shown in \autoref{fig:Preference_NASA}. 

These results show that when using PromptCanvas, 39\% of our participants felt less frustrated, while 50\% did not feel frustrated with either UI.
Additionally, 39\% perceived less temporal demand with PromptCanvas, while
39\% of participants did not feel hurried or rushed with either UI. Regarding the feeling of success, 61\% of participants felt more successful in accomplishing tasks when using PromptCanvas, and lastly, 56\% of participants reported needing less effort with PromptCanvas to accomplish their level of performance.

\section{Discussion and Future Work}

\subsection{Study Limitations}
Our study has several limitations. The small sample size ($N$ = 18) restricts generalizability. 
Future research could include a larger, more varied sample and examine how professional writers use PromptCanvas. 
Future work could also incorporate qualitative studies with thematic analysis to better understand users' experiences and how PromptCanvas supports creative writing.

\subsection{Widget Interdependence}
Dynamic widgets in PromptCanvas operate as self-contained interface objects currently. Still, changes in one widget can semantically influence others (i.e., altering a story character, such as having a son instead of a daughter, might adjust related suggestions such as names). Future iterations could try to support such semantic interdependencies.

\subsection{Supporting Creativity}
PromptCanvas significantly enhances perceived creativity support compared to traditional conversational UI by providing an interactive environment where users can explore ideas and refine outputs through dynamic widgets. This approach supports creative writing, especially for non-professional writers, by offering tools that simplify tasks and encourage experimentation with different text elements. By reducing cognitive load and frustration, PromptCanvas empowers users to express their ideas more effectively. Participants felt less hurried and more successful in their tasks, requiring fewer prompts to achieve their goals, thanks to the structured workflows enabled by dynamic widgets.

\subsection{Reducing Cognitive Load}
The NASA-TLX ratings revealed statistically significant differences in mental demand and frustration between the UIs. Participants noted feeling less annoyed, more productive, and able to complete tasks with less effort, suggesting that the dynamic interface offers a more engaging and efficient creative process. By allowing users to manipulate widgets for contextual suggestions and seamless exploration, PromptCanvas creates a supportive environment for managing creative writing tasks.

\subsection{Extending the Concept to Other Domains}
Inspired by previous research in visualization (\textit{DynaVis}~\cite{dynavis_2024}) and systems like \textit{Luminate}~\cite{suh2024luminate}, PromptCanvas demonstrates its potential for broader applications. For instance, it could extend beyond text-based creative tasks to visual content generation, where dynamic widgets might allow users to iteratively refine and customize image outputs. This capability would enable users to explore artistic styles, integrate specific elements, and adjust parameters, showcasing the versatility of dynamic widgets in diverse creative domains. The canvas-based design of PromptCanvas also allows for a broad range of widget types. Future additions could include more standard HTML elements like date-pickers, sliders, and checkboxes, similar to those in \textit{DynaVis}~\cite{dynavis_2024}, or specialized tool sets from systems like \textit{Textoshop}~\cite{massonTextoshopInteractionsInspired2024}, to allow for more targeted customization and user control.

\section{Conclusion}
In this work, we introduce PromptCanvas, using dynamic widgets as a novel solution to address the limitations of current UIs for generative AI in creative writing. Our study with 18 participants demonstrates that by incorporating customizable interactive elements, our system enhances user control, reduces cognitive load, and supports iterative exploration and the creation of a personalized design space. The findings reveal that dynamic widgets significantly improve user experience and facilitate more effective, user-centered interaction with AI. This research emphasizes the importance of user-driven customization and flexibility in unlocking the full creative potential of generative AI, leading to more meaningful and productive writing interactions.

\begin{acks}
Funded by Elitenetzwerk Bayern and the Deutsche Forschungsgemeinschaft (DFG, German Research Foundation) -- 525037874.
\end{acks}

\bibliographystyle{ACM-Reference-Format}
\bibliography{bibliography}

\appendix
\section*{Appendix}
\section{Example Scenario: Writing a Short Story}\label{app: example}

Marina is an avid reader of fiction and loves to write during her free time. She maintains a blog where she writes short stories on various topics occasionally for her readers. It has been a while since Marina wrote something for her blog. She decided to write something on Survival in the Wilderness, but she is experiencing writer's block. She decided to seek AI's help to reignite her creativity and get a starting point to carry on from there. She decides to use PromptCanvas to assist her. Below, we explain her experience with PromptCanvas.
\begin{figure*}[htb]
    \centering
    \includegraphics[width=\textwidth]{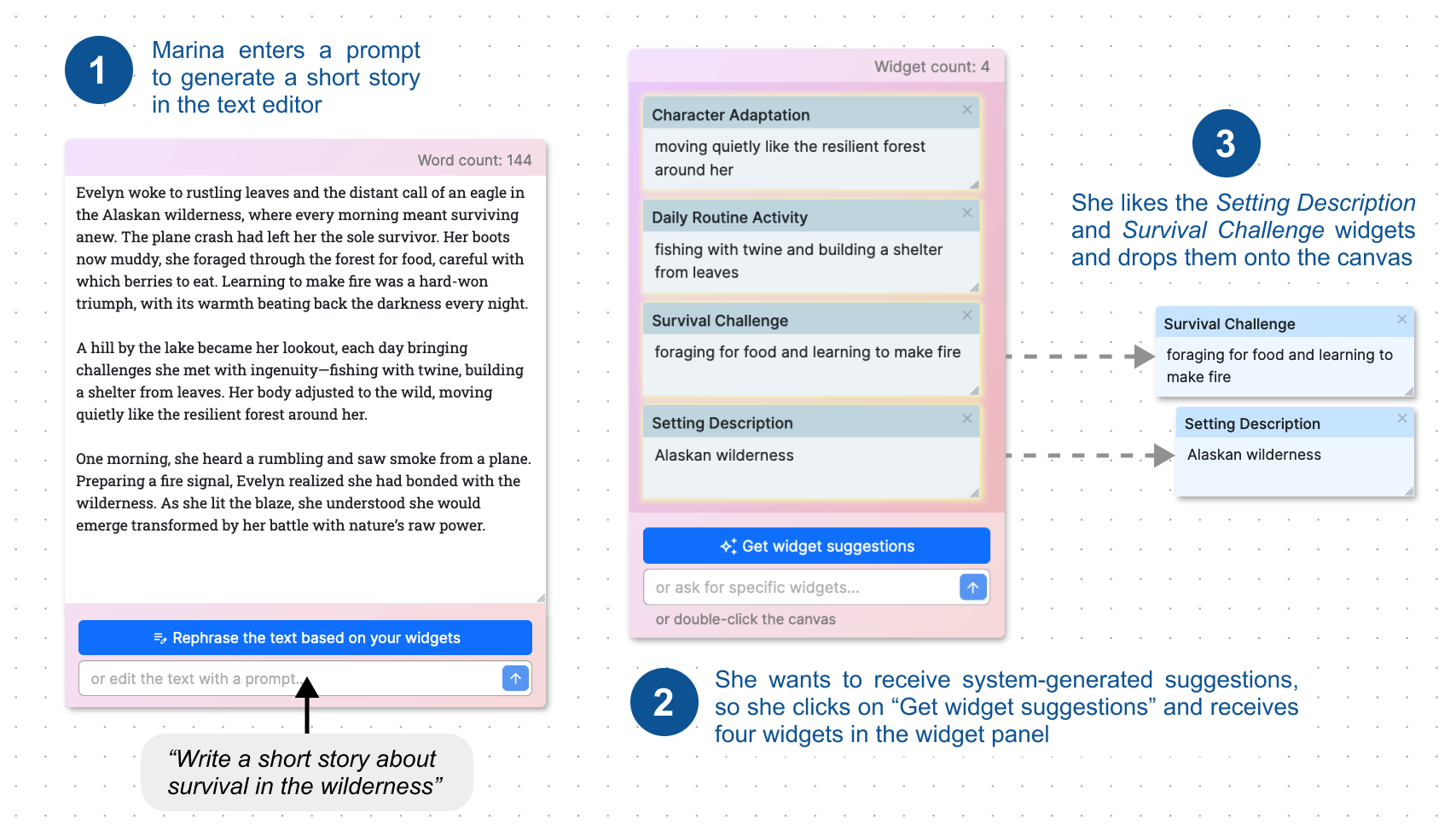}
    \caption{(1) Marina writes a prompt for the short story generation. (2) PromptCanvas generates widgets for her. (3) She chooses two widgets from the widget panel and drags them onto the canvas.}
    \label{fig:step1}
    \Description{This image depicts the text editor and widget panel, each containing exemplary generations.}
\end{figure*}

\paragraph{\textnormal{\textbf{Initial prompt/text}}}
Marina has two options to start with. She can either write directly in the text editor or generate text by writing a prompt. Opting for the latter, she initiates the process with the prompt, \textit{``Write a short story about survival in the wilderness''}, see \autoref{fig:step1}-(1). The system generates the story in the editor, but Marina wants to reiterate the story and explore more. For this, she can use the widgets. PromptCanvas provides her with three options for generating widgets: suggestions by the system, prompting to create widgets on a theme, and double-clicking on the canvas to create empty widgets.

\paragraph{\textnormal{\textbf{Widgets generated by the system}}}
Marina wants to start with the widget suggestions by the system. Therefore, she clicks on "Get widget suggestions" and receives four suggestions from the system, illustrated in \autoref{fig:step1}-(2). From there, she finds two widgets (\textit{Survival Challenge} and \textit{Setting Description}) very interesting for her story. She then drags them onto the canvas as shown in \autoref{fig:step1}-(3). She sees the colors of the widgets changing to light blue implying that the widgets are now active. 

\begin{figure*}[ht]
    \centering
    \includegraphics[width=\linewidth]{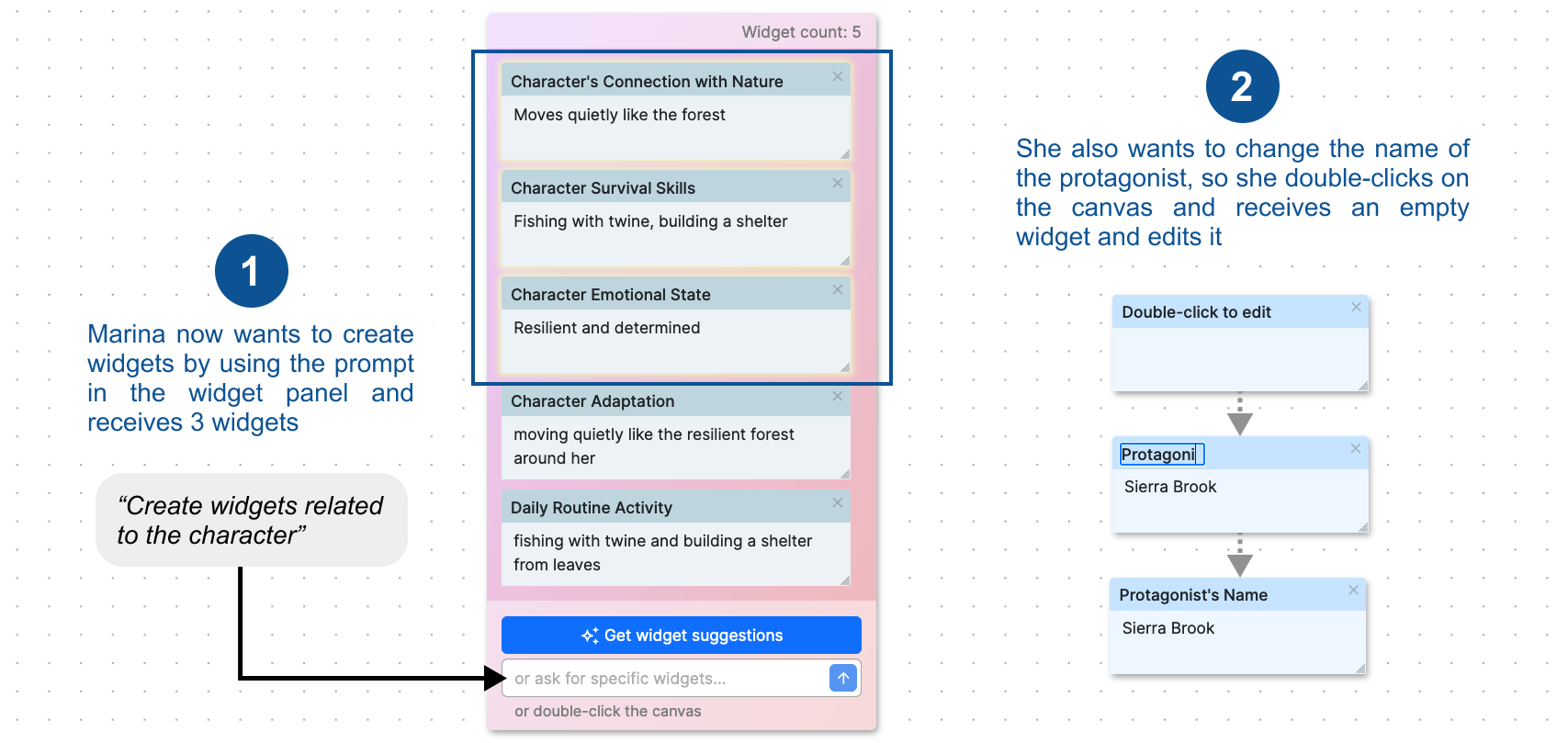}
    \caption{(1) Marina prompts in the widget panel to get more widgets. (2) She creates an empty widget on the canvas.}
    \label{fig:step2}
    \Description{This image shows the widget panel with newly generated widgets and a control widget's title being edited.}
\end{figure*}

\paragraph{\textnormal{\textbf{Prompting to get widgets}}}
As shown in \autoref{fig:step2}-(1), Marina now wants to modify some settings of the character, and to do that, she chooses to generate multiple widgets, so she decides to prompt in the widget panel \textit{``Create widgets related to the character''}. She now gets three new widgets, \textit{Character's Connection with Nature, Character Survival Skills}, and \textit{Character Emotional State} on the widget panel related to the character but with different aspects to focus on.

\paragraph{\textnormal{\textbf{Creating empty widgets}}}
Next, she aims to modify the protagonist's name and maintain the flexibility to change it as needed in the future. To ensure the generated text incorporates the name wherever necessary, she proceeds to create an empty widget on the canvas. She then edits the widget's header to \textit{Protagonist's Name} and updates the input to \textbf{Sierra Brook}, illustrated in \autoref{fig:step2}-(2).

\begin{figure*}[ht]
    \centering
    \includegraphics[width=\textwidth]{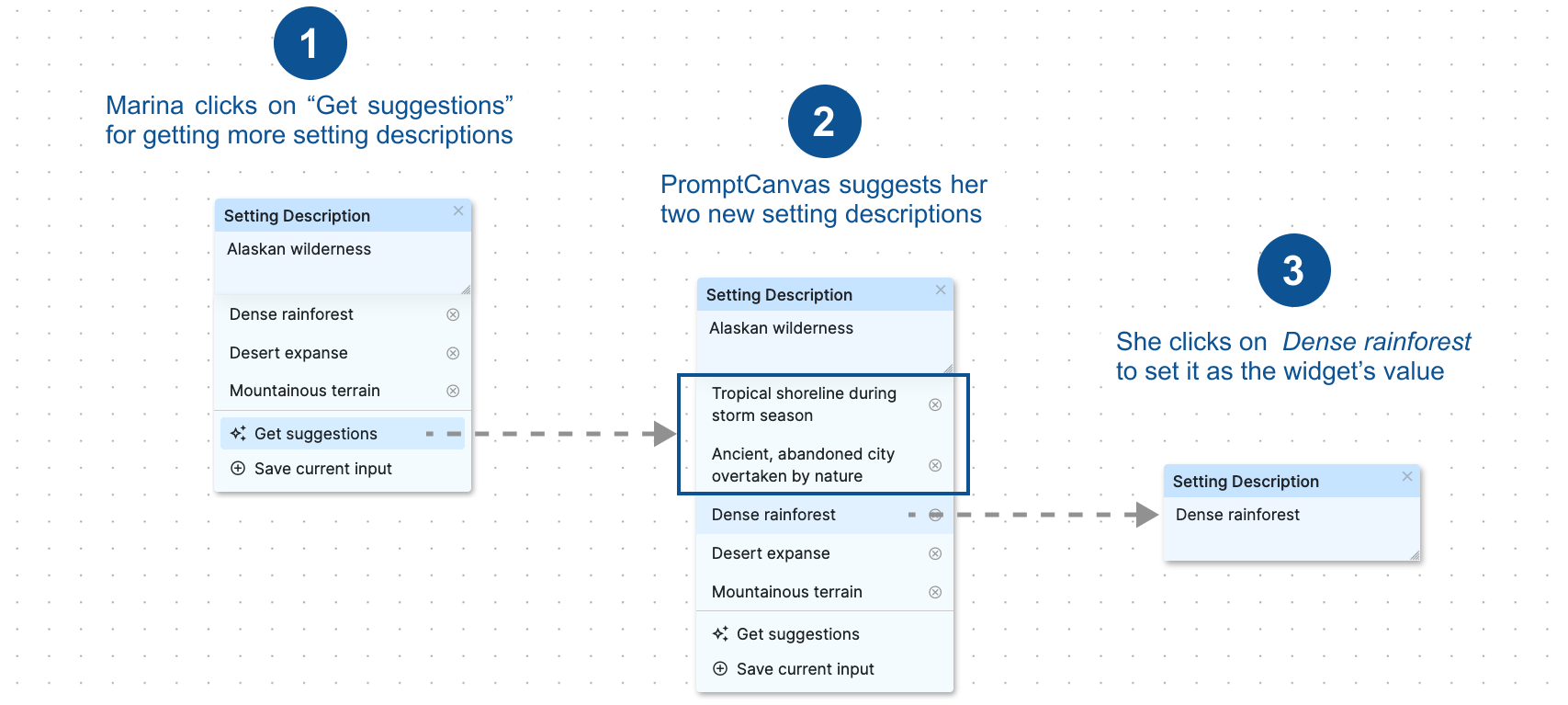}
    \caption{Marina gets more suggestions within the widget for \textit{Setting Description}.}
    \label{fig:step3}
    \Description{This image shows the suggestions generated within the control widget.}
\end{figure*}

\paragraph{\textnormal{\textbf{Suggestions within the widgets}}}
Marina does not like the current \textit{Setting Description}, so she clicks on "Get suggestions" inside the widget \textit{Setting Description}, \autoref{fig:step3}-(1). As shown in \autoref{fig:step3}-(2), PromptCanvas suggests her two new setting descriptions. After comparing the new ones with the previous suggestions, she decides to go with the \textit{Dense rainforest}, see \autoref{fig:step3}-(3).

\begin{figure*}[ht]  
    \centering
    \includegraphics[width=\textwidth]{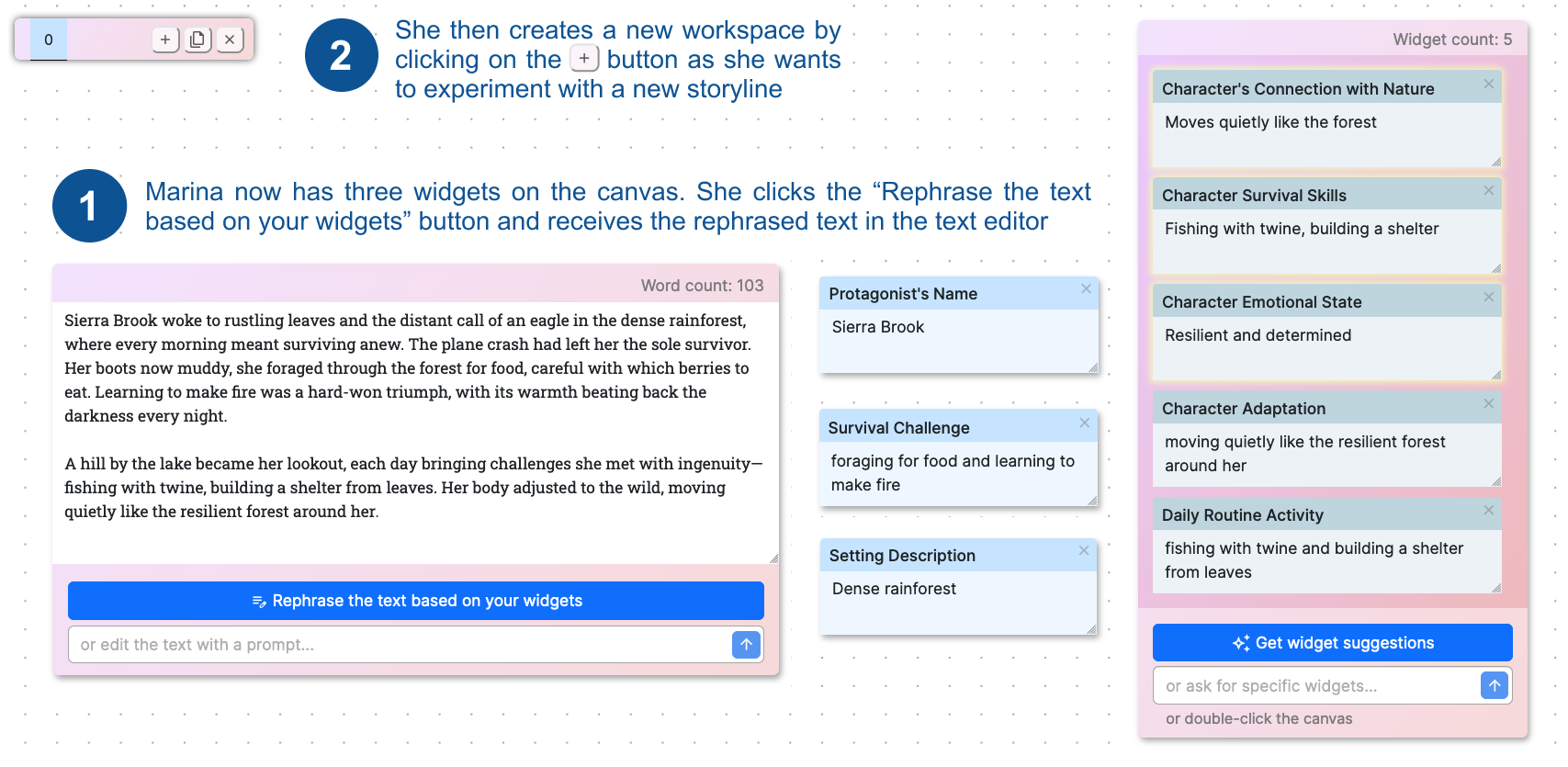}
    \caption{(1) Marina applies the widgets and receives the rephrased text. (2) She creates a new workspace using the menubar.}
    \label{fig:step4}
    \Description{This image depicts the actions taken and changes received for rephrasing the text based on control widgets.}
\end{figure*}

\paragraph{\textnormal{\textbf{Rephrasing text based on widgets}}}
As depicted in \autoref{fig:step4}-(1), Marina now has three widgets on the canvas: \textit{Setting Description, Survival Challenge, and Protagonist's Name}. She applies the widgets to the text by clicking on "Rephrase the text based on your widgets" and gets a rephrased text in the editor based on the widgets.

\paragraph{\textnormal{\textbf{Interaction with the canvas}}}
After getting the rephrased text, Marina feels like she wants to try out a different storyline and, therefore, creates a new canvas by clicking on the \textbf{+} button on the menu bar, illustrated in \autoref{fig:step4}-(2).

The workflow above allows Marina to experiment with different ideas. It highlights the dynamic interaction between creativity and structure which empowers Marina to overcome her writer's block and shape her story effectively. This approach, however, showcases only one of the many diverse possibilities of how a user can use PromptCanvas.

\section{Apparatus}
\begin{figure*}[ht!]
    \centering\includegraphics[width=\textwidth, trim={0cm 0cm 0cm 0cm},clip]{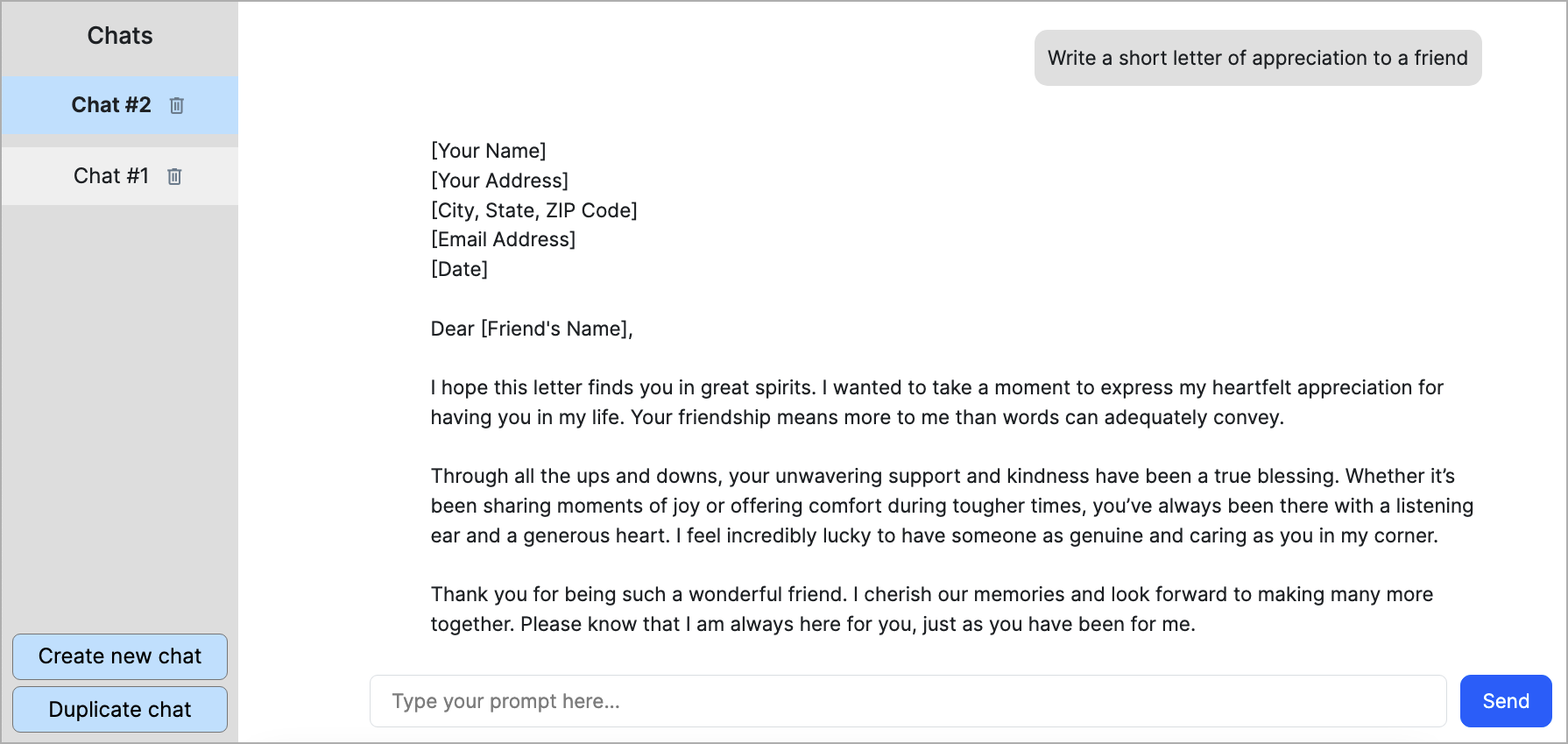}
    \caption{The baseline conversational UI.}
    \label{fig:static_UI}
    \Description{This image shows the baseline conversational UI.}
\end{figure*}

Our user study included two conditions. The baseline condition used a conversational user interface as illustrated in \autoref{fig:static_UI}, while the experimental condition used PromptCanvas. 
The baseline system was designed according to the design and interaction principles of ChatGPT. We provided a solid user experience without introducing untested features that could have affected the study. 
To generate the responses, we used the same OpenAI model (gpt-4o-2024-08-06) in both conditions.
On the left side of the UI is a sidebar in which all chat instances are listed to be selected or deleted, and buttons for creating a new or duplicating the currently selected chat. Selected chats are displayed chronologically in the main component by listing all user and assistant messages. Below the chat messages is a text input for entering new user messages. Responses are received in a stream and displayed as received, with words and sentences gradually appearing as if they were typed. While hovering over a message, a small icon appears below for easily copying the message's content. There is also an edit icon for user messages to alter the message and reset the chat to that point.

\section{User Study Details} \label{app: user study}
\subsection{Participants} \label{app: participants}
All participants had previous experience with (creative) writing. Specifically, 15 had experience in writing emails, 13 in writing letters, 9 in writing articles, 7 in writing stories, 6 in writing blogs, 5 in writing poems, 4 in writing how-to guides, 3 in writing product reviews, 2 in copywriting, 1 in song composition, 1 in character development, and 1 in writing travel guides. 

Participants also used AI tools for various writing tasks. 12 participants used these tools for editing and proofreading, followed by 9 who used them for idea generation and 8 for content expansion. 6 participants used AI tools for descriptive writing, and another 6 for creating different versions of their writing. Additionally, 3 participants used AI tools for creative writing, while only one used them for translation and another for coding. Apart from 2 participants who had never used AI tools for any writing task, everyone else had experience using them for various writing purposes.

Regarding concrete AI tools, 15 participants used ChatGPT, 6 Quillbot, 5 Bard (known as Gemini now), and 3 Claude. Additionally, 3 participants had never used any tools, 2 used DALL-E, 1 Perplexity AI, 1 Stable Diffusion, and 1 Grammarly. 

Regarding the frequency of AI writing tool usage, 7 participants used them daily, 5 weekly, 2 monthly, 2 rarely, and 2 had never used them. For the study, 12 participants used a laptop, and 6 used a big screen (e.g., an external monitor). 11 participants used a touchpad, while 7 used a mouse.

\subsection{Writing Tasks}
\label{app:writingTasks}

\begin{table}[!htb]
\caption{Topics for the writing tasks in the user study.}
\begin{tabularx}{\linewidth}{p{2cm}X}
\toprule
   Writing Tasks  & Topic \\ 
   \midrule
    Email or Letter  & Professional
        \begin{itemize}
            \item Resignation letter
            \item Motivation letter for job application
            \item Recommendation letter
            \item Request for promotion
        \end{itemize}\\
        & Personal
        \begin{itemize}
            \item Condolences
            \item Updates on Life
            \item Friendship and Appreciation
            \item Celebrations and Milestones
        \end{itemize}\\ 
    \midrule
    Short story & Survival in the Wilderness\\ 
        & AI robots\\ 
        & Time travel\\ 
        & Life after Death\\ 
        & Family Secrets\\ 
        & Utopia / Dystopia\\ 
        & Fable \\ 

\bottomrule

\end{tabularx}
\Description{This table shows the topics for the writing tasks in the user study.}
\label{tab:tasks}
\end{table}

\end{document}